\documentclass[prl,superscriptaddress,twocolumn,showpacs,
amsfonts,amsmath,amssymb]{revtex4}
\usepackage{graphicx}% Include figure files
\usepackage{bm}% bold math

\newcommand{\y}{Y}

\catcode`\@=11
\def \matrix #1 {\left(\begin{array}{cc} #1 \end{array}\right)}

\def \tr {\mathop{\rm tr}\nolimits}

\newcommand{\as}{\ifmmode\alpha_{\rm s}\else{$\alpha_{\rm s}$}\fi}
\newcommand{\asbar}{\ifmmode\bar{\alpha}_{\rm s}\else{$\bar{\alpha}_{\rm s}$}\fi}
\newbox\lett\newdimen\lheight\newdimen\lwidth
\def\ontop#1#2{\setbox\lett=\hbox{#2}\lheight\ht\lett
\multiply\lheight by 12 \divide\lheight by 10\relax%
\lwidth\wd\lett \multiply\lwidth by 8 \divide\lwidth by 10\relax #2\kern-\lwidth%
\raise\lheight\hbox{{$\scriptstyle #1$}}\kern.1ex}

\begin{document}

%\preprint{RUB-TPII-03/05}

\title{Extension of the Color Glass Condensate Approach to Diffractive
Reactions}
\author{Martin Hentschinski, Heribert Weigert, and Andreas Sch\"afer}
\affiliation{Institut f\"ur Theoretische Physik,
             Universit\"at Regensburg,
             D-93040 Regensburg, Germany}

\date{\today}% It is always \today, today,

\begin{abstract}
  We present an evolution equation for the Bjorken $x$ dependence of
  diffractive dissociation on hadrons and nuclei at high energies. We extend
  the formulation of Kovchegov and Levin by relaxing the factorization
  assumption used there. The formulation is based on a technique used by
  Weigert to describe interjet energy flow.  The method can be naturally
  extended to other exclusive observables.
\end{abstract}
\maketitle

QCD at very high parton densities is one of the most active frontiers both in
high-energy and nuclear physics and one of the topics where both fields
clearly profit from close collaboration.  With the advent of the LHC in 2007
this topic will further gain importance. Both the search for new physics in
proton-proton collisions and the investigation of high energy medium effects
in heavy ion collisions require a solid understanding of multiple gluonic
interactions (at the very least for the analysis of backgrounds).  Presently,
the rapid output of precise experimental data at RHIC, where the same effects
should be present, though less pronounced, provides the main driving force
behind new theoretical developments.  One of the theoretically most attractive
approaches is known under the name of color glass condensate
(CGC)~\cite{heribert} and one of its main elements is the JIMWLK-equation
describing the evolution of characteristic quantities with the squared cm
energy $s$~\cite{JIMWLK}.  Our paper is based on this approach.

At high energies, as reached in RHIC and LHC experiments, most QCD observables
receive strong contributions from multiple soft gluon emission and multiple
interactions of ``hard'' particles with soft gluons present in the event.  A
reliable and transparent method to resum the effects of these soft gluons on
the hard leading particles can be formulated by using gauge links $U(x;y)={\sf
  P} \exp - i g\int_y^x dz.A(z)$ where the trajectories (from $y$ to $x$)
represent the quasiclassical paths of the hard particles while the soft gluons
appear in the exponent. Previous work has focused on inclusive reactions.
Here we demonstrate how to extend this program to exclusive reactions and work
out the example of diffractive dissociation, where we can compare to a known
limiting case~\cite{kl} that emerges if we use a factorization assumption as
in the reduction of the JIMWLK to the Balitsky-Kovchegov (BK)~\cite{ian, BK}
equation.

As an example let us recall that e.g. the total cross section of deep
inelastic scattering of a virtual photon on a nuclear target can be
written in terms of the $U$s as:
\begin{equation}
 \label{equ-1}
 \begin{split}
 \sigma_{\text{DIS}}(\y,Q^2)
&= 
\int_0^{{}_1}
% \int_0 \! {}^{{}^{{}^1}}
 \!\! d \alpha
 \!\int\!\! d^2 \bm{r}~|\psi^2|({\bm{r}}^2 \alpha (1-\alpha)Q^2)
\\
 & \qquad \qquad \quad % \qquad \cdot \
 \!\int\!\! d^2 \bm{b} \
 \left \langle %\frac
 {{\rm tr}[1-U_{\bm{x}}U_{\bm{y}}^{\dagger}]}/{N_c} \right
 \rangle_\y 
 \end{split}
\end{equation}
where $|\psi^2|({\bm{r}}^2 \alpha (1-\alpha) Q^2)$ is the probability of a
photon to split into a quark-antiquark pair of size $\bm{r}=\bm{x}-\bm{y}$,
carrying longitudinal momentum fractions $\alpha$ and $1-\alpha$,
respectively.  The remaining integral over the impact parameter
$\bm{b}=(\bm{x}+\bm{y})/2$ yields the cross section of a $q\Bar q$ dipole of
size $\bm{r}$. The rapidity $\y = \ln 1/x$ is taken to be large. The gauge
links $U_{\bm{x}}$ and $U_{\bm{y}}^\dagger$ represent the leading hard quark
and antiquark within the virtual photon wave function. They propagate at fixed
transverse coordinates $\bm{x}$ and $\bm{y}$ along straight lines from
$z^-=-\infty$ to $z^-=\infty$:
\begin{align}
  \label{equ2}
U_{\bm{x}} = &\ 
{\sf P} \exp - i g\int\limits_{-\infty}^{\infty}\!\!dz^-
b^+ (z^-,\bm{x},x^+=0)
\ .
\end{align}
In \eqref{equ2} we have anticipated that the hard particles interact, to
leading order, only with $b^+$, the soft %, frozen 
component of the gauge field
with rapidities below some $\y_0$. Generically $A^\mu(x)= \ b^\mu(x)+\delta
A^\mu(x)$ with $b^+ = \delta(x^-)\beta(\bm{x})$ and $ b^-=\bm{b}=0$; $\delta
A^\mu$ denotes all hard fluctuations.

The averaging indicated in~\eqref{equ-1} is over the soft fields $b^+$ and
represents all the interactions with the target through gluons softer than the
original $q\Bar q$. As such it contains nonperturbative information that can
not directly be calculated.  Since this decomposition into hard and soft modes
is rapidity dependent, the dipole cross section turns rapidity dependent as
well.  By considering hard corrections $\delta A$ one can systematically
calculate the $\y$ dependence and find RG equations for the dipole cross
section. A direct approach leads to an infinite hierarchy of equations, the
Balitsky hierarchy~\cite{ian}, which can only be solved after truncation. A
more compact formulation in terms of a single diffusion equation can be given
in a functional language. To this end one parametrizes the lack of knowledge
about the averaging procedure by using a functional $\Hat Z_\y[U]$, which
takes on the meaning of a statistical distribution function:
\begin{equation}
  \label{eq:Zdef-DIS}
  \langle \ldots \rangle_\y := \int \Hat D[U] \ldots \Hat Z_\y[U]
\ .
\end{equation}
$\Hat D[U]$ is a Haar measure that is normalized to 1. The $s$-, respectively
$Y$-, evolution for the dipole cross section and all the other more
complicated correlators in the Balitsky hierarchy is then given by the JIMWLK
equation, which governs the evolution of the functional weight $\hat{Z}$:
\begin{equation}
\label{eq:JIMWLK}
\partial_\y \hat Z[U]_Y = 
- H_{\text{JIMWLK}}
~\hat Z[U]_\y \ .
\end{equation}
\eqref{eq:JIMWLK} describes a Fokker-Planck type diffusion problem in
a functional context. The JIMWLK Hamiltonian 
\begin{equation}
  \label{eq:H_JIMWLK}
  \begin{split}
  H_{\text{JIMWLK}}
  =-\frac{\alpha_s}{2\pi^2}{\cal K}_{\bm{x z y}}
   & \Big(
 U^{a b}_{\bm{z}} (
    i\nabla^a_{\bm{x}} i\Bar\nabla^b_{\bm{y}}
    + i\Bar\nabla^a_{\bm{x}} i\nabla^b_{\bm{y}}
    )
 \\  & \hspace{.5cm}
+ i\nabla^a_{\bm{x}} i\nabla^a_{\bm{y}} + i\Bar\nabla^a_{\bm{x}} i\Bar\nabla^a_{\bm{y}}
    \Big)    
  \end{split}
\end{equation}
(integration over repeated transverse coordinates is implied here and below)
contains a real emission part proportional to a new adjoint Wilson line $
U^{ab}_{\bm{z}}$ that signals the appearance of a new gluon in the final state
and virtual corrections that guarantee finiteness of the evolution equation.
The remaining ingredients are the kernel ${\cal K}_{\bm{x z y}} =\ 
\frac{(\bm{x}-\bm{z}){\cdot}(\bm{z}-\bm{y})}
{(\bm{x}-\bm{z})^2(\bm{z}-\bm{y})^2}$ and functional derivatives
$i\nabla^a_{\bm{x}}$ that respect the group valued nature of the $U$-fields: $
i\nabla^a_{\bm{x}} := -[U_{\bm{x}} t^a]_{i j} {\delta}/{\delta U_{\bm{x},i j}}
$ corresponds to the left invariant vector field on the group manifold, while
its right invariant counterpart is given by $ i\Bar\nabla^a_{\bm{x}} := [t^a
U_{\bm{x}} ]_{i j} {\delta}/{\delta U_{\bm{x},i j}} = - U_{\bm{x}}^{b a}
i\nabla^b_{\bm{x}} $.

To prepare for the treatment of non-inclusive observables, we will now
sketch how to recover this evolution equation from the underlying real
emission amplitudes. The treatment parallels that of jet observables
in~\cite{jets}. To facilitate this construction, let
us introduce Wilson lines $U_{\y, \bm{x}}$ that are (slightly) tilted
w.r.t. the lightcone. The derivation is then based on the observation
that the whole cloud of $\y$ ordered real gluons accompanying any
given number of hard partons that can be characterized as a product of
Wilson lines $U_{\y_1, \bm{x_1}}^{(\dagger)} \cdots U_{\y_n,
  \bm{x_n}}^{(\dagger)}$ can be generated by the application of a
single operator
\begin{align}
  \label{eq:real_emission}
  \begin{split}
  {\sf U}[U,\xi] =   {\sf P}_{\y_2} \exp\Big[
  i\!\! \int\!\! &\ d\y_1  d\y_2 
 \theta(\y_1\!-\!\y_2)
\\ &
  J_{\bm{x z}}^i
   U_{\y_2, \bm{z}}^{ab} \xi^{b, i}_{\y_2, \bm{z}}  
  i\Bar \nabla^a_{\y_1, \bm{x}}
  \Big]    
  \end{split}
\end{align}
with $J_{\bm{x z}}^i := \tfrac{g}{4 \pi^2}
\tfrac{({\bm{x}}-{\bm{z}})^i}{({\bm{x}}-{\bm{z}})^2}$ the eikonal current in
transverse coordinate space. The $\xi$ fields represent the gluonic final
states.  The derivatives now act as $ i\nabla^a_{\y, \bm{x}} := -[U_{\y,
  \bm{x}} t^a]_{i j} {\delta}/{\delta U_{\y, \bm{x},i j}} $; $\y$-ordering is
such that the hardest gluon is rightmost. This ensures that gluons can only
emit softer ones.  For DIS, the hard seed consists of a $q\Bar q$ pair
represented by a product of Wilson lines $U_{\y, \bm{x}} U_{\y,
  \bm{y}}^\dagger$ at projectile rapidities.  [The external color indices are
those of the amplitude.]  Diagrammatically we get
\begin{equation}
  \label{eq:JIMWLK-real-diags}
  {\sf U}[U,\xi] U_{\y, \bm{x}} U_{\y, \bm{y}}^\dagger = \sum\limits_{
      \begin{minipage}{2cm}\centering
        \tiny
        $\#$ of gluons \& allowed insertions
      \end{minipage}}  
\hspace{-.5cm}\parbox{2.5cm}{\includegraphics[height=2cm]{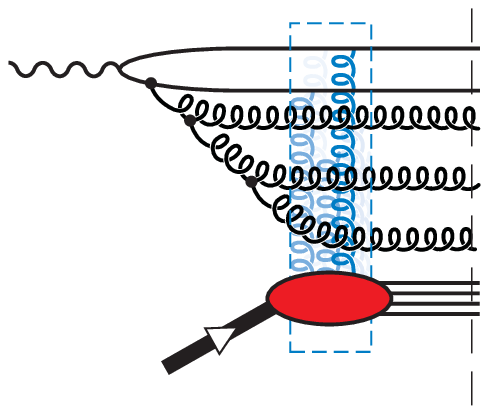}}
\end{equation}
where the vertical dashed line denotes the final state, where each line ends
in a factor $\xi$.  These stand for explicitly resolved partons in an interval
$[\y_0,\y]$ over which we follow the logarithmically enhanced contributions.
The remaining gluons indicate soft interactions with the target below $\y_0$,
which build up the $U$ fields. They correspond to the initial condition of
the evolution process. To recover the real emission part of the dipole cross
section, we need to square this amplitude and integrate over phase space for
the resolved gluons. The average over the soft, unresolved gluons is 
done separately in amplitude and complex conjugate amplitude
\cite{kovwied}: %. To this end, 
we distinguish corresponding eikonal factors $U$ and $\Bar U$. The
average over the resolved final states can be made explicit by averaging over
the final state variables $\xi$ with a Gaussian
weight.  %Such an average over any 
For an arbitrary functional $F[\xi]$ this is expressed
as~\cite{jets}
\begin{align}
  \label{eq:gauss_average_legendre}
\langle F[\xi] \rangle_\xi = 
    \exp\left\{
{ - \frac{1}{2} \frac{\delta}{i \delta \xi} M  \frac{\delta}{i \delta \xi} }
\right\} 
F[\xi]\bigg|_{\xi =0}.
\end{align}
The $\xi$-correlator ${M}^{a,i}_{\y_1 \bm{u}} {}^{b,j}_{\y_2
  \bm{v}}=\langle \xi^{a, i}_{\y_1 \bm{u}}  \xi^{b, j}_{\y_2
  \bm{v}}\rangle$, appropriately normalized, is given by
\begin{equation}
  \label{eq:xi-correlator}
{M}^{a,i}_{\y_1 \bm{u}} {}^{b,j}_{\y_2 \bm{v}}%(\y)
  =
4\pi\delta^{ab}\delta^{ij} \delta^{(2)}_{\bm{u},\bm{v}}\delta_{\y_1,\y_2}
\theta(\y \!-\! \y_1)\theta(\y_1 - \y_0)
\ .
\end{equation}
$M$ is diagonal in coordinates and rapidities and restricted to the resolved
evolution interval $[\y_0,\y]$. In the exponent
of~\eqref{eq:gauss_average_legendre} integration and summation over all
indices is understood.  The resolved contribution of real emissions to the
dipole cross section is thus obtained by the Gaussian average
\eqref{eq:gauss_average_legendre} over the functional
\begin{equation}
  \label{eq:dresseddipole}
G^{\text{real}}[\xi]
= 
       {\sf U}[U, \xi] {\sf U}[\bar{U}, \xi] 
       \frac{
       \tr[1- (U \Bar U^\dagger)_{\y, \bm{x}} 
       (U \Bar U^\dagger)^\dagger_{\y,\bm{y}}]
       }{N_c}
\end{equation}
where $\hat{N}_{\bm{x y}}^F = \tr[1- (U \Bar U^\dagger)_{\y,\bm{x}} (U \Bar
U^\dagger)^\dagger_{\y,\bm{y}}]/N_c $ is the dipole operator of the total
cross-section. No matter how the average over the non-resolved modes below
$\y_0$ is achieved, the evolution of the complete real emission part is
determined by the $\y$ dependence of the resolved contributions:
\begin{align}
  \label{eq:evolv_GvonXi_IV}
\partial_\y \langle G^{\text{real}}[\xi] \rangle_\xi
=&
         -\frac{  e^{ 
             - \frac{1}{2} \frac{\delta}{i \delta \xi} M  \frac{\delta}{i
               \delta \xi} 
           }
         }{2}     
         \frac{\delta}{i \delta \xi} \partial_\y M(\y)  
         \frac{\delta}{i \delta \xi}        
         G[\xi]\bigg|_{\xi =0}
\notag \\
=
   \langle  
 {\sf U}[U, \xi] 
{\sf U}[\bar{U}, \xi]  
 &
\frac{\alpha_s}{\pi^2} \mathcal{K}_{\bm{x z y}} 
(U\bar{U}^\dag)^{ab}_{\y, \bm{z}} 
i\Bar \nabla^a_{U_{\y,\bm{x}}}    i\Bar \nabla^b_{\bar{U}_{\y,\bm{y}}} 
\!\!   
\hat{N}_{\bm{x y}}^F    
\rangle_\xi
\ .
\end{align}
In this equation everything outside the shower operators only contains $U$
factors at the upper $\y$ limit. This allows to recast both of these averages
in terms of averages over Wilson lines at this highest rapidity: one can set
\begin{align}
  \label{eq:average-change}
  \langle ... \rangle_\y = \langle {\sf U} [U, \xi] {\sf U}[\bar{U}, \xi] ...    \rangle_{\xi,\text{soft}}
=\!\!
\int\!\! \Hat D[U] \Hat D[\Bar U] 
    ... \Hat Z_\y[U,\Bar U]
\end{align}
We may drop the now unnecessary $\y$-label on the Wilson lines.  This result,
as all inclusive quantities, only depends on products $U \Bar U^\dagger$ in
the hard operators appearing in~(\ref{eq:evolv_GvonXi_IV}) and thus also in
the weight $\hat Z$. By a redefinition $U \Bar U^\dagger\to U$, the
integration over $\Bar U$ then reduces to a factor of 1 and one is back
at~\eqref{eq:Zdef-DIS}. (\ref{eq:evolv_GvonXi_IV}) then leads to the
contribution of real emissions to the evolution equation
\begin{align}
   \label{eq:evol-real}
      (\partial_\y \hat{Z}[U ]  )^{\text{{real}}}
\!\! =
  \frac{\alpha_s}{\pi^2} \mathcal{K}_{\bm{x z y}} 
    \, U^{a b}_{\bm{z}}\, 
  i\Bar \nabla^a_{U_{\bm{x}}}    
  i \nabla^b_{U_{\bm{y}}}  
  \hat{Z}[U]
\ .
\end{align}
Inserting virtual corrections by the requirement of real-virtual cancellation
in absence of interaction, one recovers the JIMWLK evolution as stated
in~(\ref{eq:JIMWLK}, \ref{eq:H_JIMWLK}).  Exclusive quantities on the other
hand will depend separately on $U$ and $\Bar U$ and require to keep both
fields in $\Hat Z$ along with more complicated evolution equations.

\begin{figure}[htb]
  \centering
\includegraphics[height=2cm]{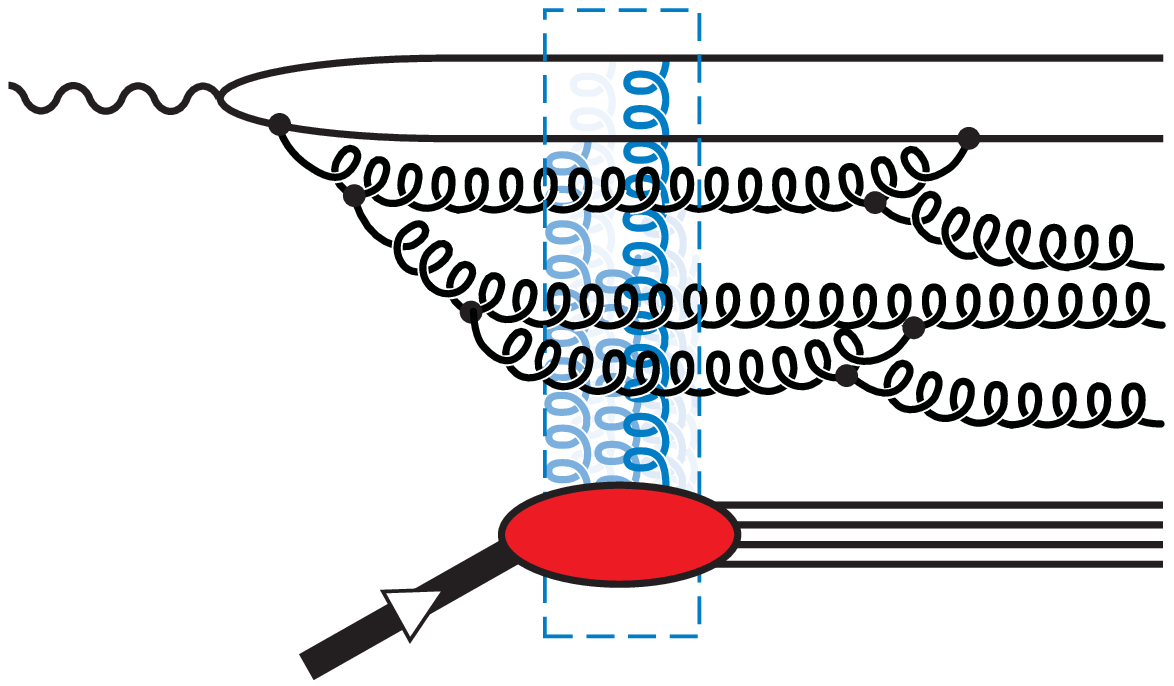}
\hspace{.5cm}
\includegraphics[height=2cm]{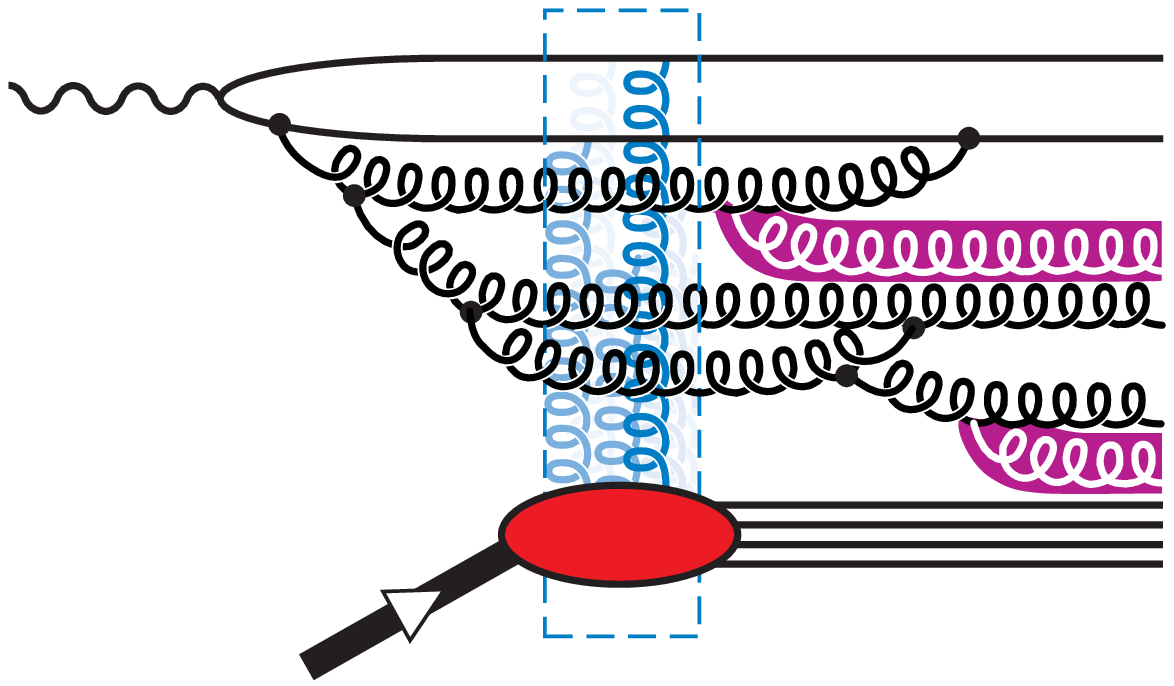}
        \caption{Generic diagrams for exclusive processes with final
          state interactions. In the diagrams, rapidity of gluons increases
          both vertically, in the final state, and horizontally, with the
          distance of their emission vertex to the target: To leading
          logarithmic accuracy, ordering in $\y$ coincides with ordering in
          $z^-$ towards the interaction region.  Consequently, emissions into
          the final state after the interaction do not iterate: lines marked
          in the graph to the right are suppressed.}
  \label{fig:generic-exclusive}
\end{figure}

Since exclusive quantities are characterized by specific restrictions on the
phase space of produced gluons the physically most transparent derivation of a
corresponding evolution equation is built on a systematic construction of the
contributing real emission amplitudes.  The first modification clearly
concerns the $\xi$-correlator $M$ used to implement the phase space integrals.
Diffractive dissociation, which corresponds to a rapidity gap on the side of
the target, requires a factor $u(k)=\theta(\y_k-\y_{\text{gap}})$ for each
final state gluon with momentum $k$ . (The gap rapidity $Y_{\text{gap}}$ is
assumed to lie in the resolved range.)  The major change, however, results
from the appearance of additional diagrams that disappear in the inclusive
result through complete real virtual cancellation.  While for JIMWLK it is
sufficient to consider branching processes that occur before the interaction
with the Lorentz contracted target, exclusive observables like diffractive
dissociation will receive contributions from reabsorption and production in
the final state, i.e. after the interaction with the target as shown in
Fig.\ref{fig:generic-exclusive}.
Reabsorption of a gluon after the interaction in the amplitude takes a form
similar to a virtual correction in the JIMWLK case, but contains the soft
interaction with the target, i.e. a factor $U$ per hard particle. Technically,
the necessary diagrams can be constructed by introducing a ``three time
formalism'' in which we distinguish $z^-=-\infty, =0$ and $=+\infty$ as the
times at which the initial hard particles are created, the interaction takes
place and the final state is formed respectively. The transition amplitude
from $z^-=-\infty$ to $+\infty$ is then created in two steps: we use a shower
operator to create gluons before the interaction but anticipate that some of
them directly reach the final state while others will be reabsorbed after the
interaction.  In order to also generate the final state contributions with a
shower operator, we introduce an auxiliary Gaussian ``noise'' $\Xi$ with the
same average and correlator as in~\eqref{eq:gauss_average_legendre}
and~\eqref{eq:xi-correlator}. Furthermore we artificially split the $U$
factors of the interaction region into two Wilson lines $ W$ and $V^\dagger$
according to $U= W V^\dagger$. (One may think of them as Wilson lines
extending over the intervals $[-\infty,0]$ and $[0,\infty]$, respectively,
they will disappear in the final
result.) We then obtain the full set of
diagrams:% according to
  \begin{align}
    \label{eq:diffdiss-amplitude}
 &     \langle {\sf U}_{\text{f}}[\Xi,\xi]{\sf U}_{\text{i}}[\Xi] \
     \parbox{1.8cm}{\includegraphics[width=1.8cm]{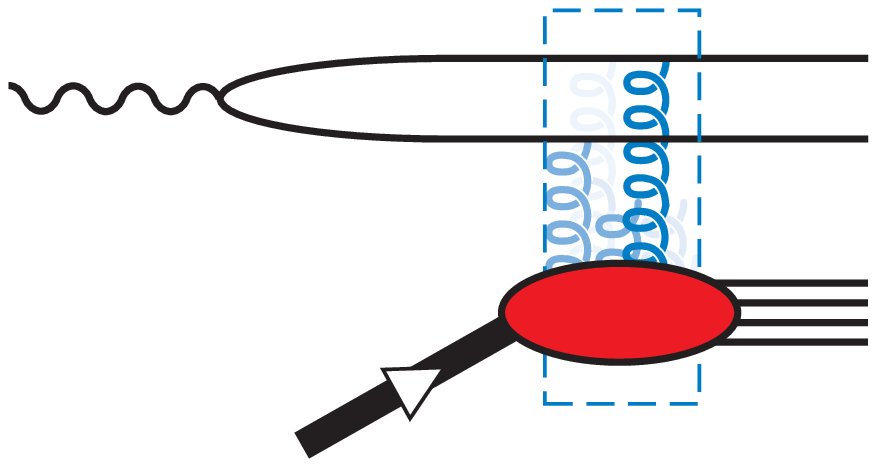}}\
     \rangle_\Xi 
\\ \notag 
&
=     \langle {\sf U}_{\text{f}}[\Xi, \xi] 
\sum \!\!
 \parbox{2.5cm}{\includegraphics[width=2.5cm]{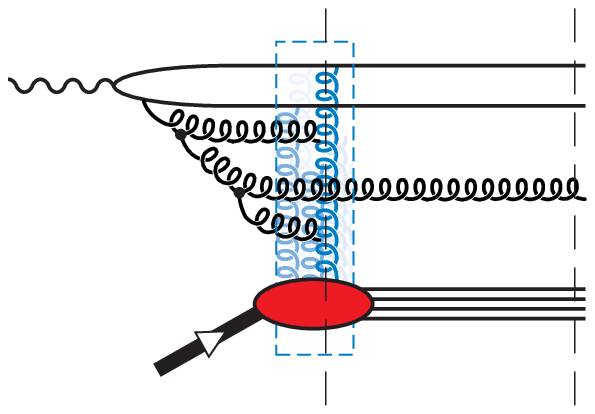}}
 \rangle_\Xi
=\sum \!\!
 \parbox{2.5cm}{\includegraphics[width=2.5cm]{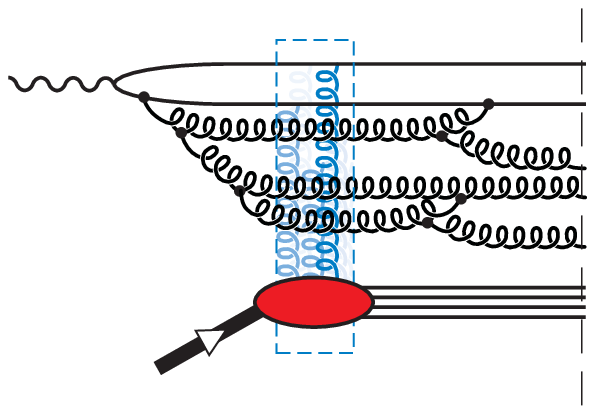}} 
\end{align}
where the sum is over the number of gluons and allowed insertions.  The dashed
line through the interaction region represents the auxiliary split of the
Wilson lines into $W$ and $V^\dag$ with accompanying $\Xi$ factors.  The
shower operators are given by
\begin{subequations}
\begin{align}
  \label{eq:Uforamplitude}
  {\sf U}_{\text{i}}& [\Xi,\xi]
=  
 {{ \sf P}}_{\y_2} \exp\Big[ i 
 \int  d\y_1d\y_2\, \theta(\y_1\!-\!\y_2) J^i_{\bm{x z}}
\notag \\
& ( 
 W^{ab}_{\y_2, \bm{z}}\Xi^{b,i}_{\y_2, \bm{z}}  
 +
 ({WV^\dag})^{ab}_{\y_2, \bm{z}}  \xi^{b,i}_{\y_2, \bm{z}}
 ) 
  i\Bar\nabla^a_{W_{\y_1\bm{x}}}
 \Big]
  \\
 {\sf U}_{\text{f}}& [\Xi,\xi]
=  
 {{\sf P}}_{\y_2}  \exp\Big[i 
 \int \!\!  d\y_1d \y_2\, \theta(\y_1\!-\!\y_2)     
J^i_{\bm{x z}} 
\notag \\
&\qquad 
( 
 V^{ab}_{\y_2, \bm{z}}\Xi^{b,i}_{\y_2, \bm{z}} 
  +    \xi^{a,i}_{\y_2, \bm{z}}) 
 i\Bar \nabla^a_{V_{ \y_1, \bm{y}}} 
 \Big] 
\ .
\end{align}  
\end{subequations} 
Eventually combining the above expression for the amplitude with the
corresponding expression for the complex conjugate amplitude and
differentiating w.r.t. $\y$ yields all real emission contributions to the
evolution Hamiltonian as well as the interacting virtual ones. One still
misses virtual lines that do not cross the interaction regions.  These are
again reconstructed on the level of the evolution equation. We obtain the full
Hamiltonian:
\begin{align}
  \label{eq:full-H}
  H &= u(k) H_r + H_v  + H_{\bar{v}} 
\end{align}
where the real gluonic corrections  are produced by
\begin{align*}
  H_r &=
- \frac{\alpha_s}{\pi^2} {\cal K}_{\bm{x z y}}\big(
   U^{ab}_{\bm{z}} i\Bar\nabla^a_{U_{\bm{x}}} i \nabla^b_{\bar{U}_{\bm{y}}} 
+ \bar{U}^{ab}_{\bm{z}} i \Bar\nabla^a_{\bar{U}_{\bm{x}}}  i \nabla^b_{{U}_{\bm{y}}}  
\notag \\
& \qquad \qquad \qquad 
+ (U\bar{U}^\dag )^{ab}_{\bm{z}}  i\Bar\nabla^a_{{U}_{\bm{x}}} i\Bar\nabla^b_{\bar{U}_{\bm{y}}} 
+  i \nabla^a_{{U}_{\bm{x}}} i \nabla^a_{\bar{U}_{\bm{y}}} \big)
\ .
\end{align*}
The remaining terms  correspond to virtual corrections in
amplitude and complex conjugate amplitude respectively
\begin{align*}
        H_v \!\!=&
 \frac{-\alpha_s}{2\pi^2} {\cal K}_{\bm{x z y}} ( 
i\nabla^a_{U_{\bm{x}}} i\nabla^a_{U_{\bm{y}}}\!\! 
+ \!\! i\Bar\nabla^a_{U_{\bm{x}}} i\Bar\nabla^a_{U_{\bm{y}}  } 
\!\!+\!\!
2U^{ab}_{\bm{z}}  i\Bar\nabla^a_{U_{\bm{x}}} i\nabla^b_{U_{\bm{y}}}
)
\notag \\
 H_{\bar{v}} \!\!=&   
\frac{-\alpha_s}{2\pi^2} {\cal K}_{\bm{x z y}} ( 
i\nabla^a_{\bar{U}_{\bm{x}}} i\nabla^a_{\bar{U}_{\bm{y}}} 
\!\!+\!\!  
i\Bar \nabla^a_{\bar{U}_{\bm{x}}} i\Bar \nabla^a_{\bar{U}_{\bm{y}}  } 
\!\! +\!\!
2 \bar{U}^{ab}_{\bm{z}}  
i\Bar\nabla^a_{\bar{U}_{\bm{x}}} i \nabla^b_{\bar{U}_{\bm{y}}}
)
\ .
\end{align*}
(The last terms in these expressions are the interacting parts.)  The
evolution equation parallels~(\ref{eq:JIMWLK}), with $\Hat Z$ replaced by
$\Hat Z[U,\Bar U]$. Note that $H_v$ and $H_{\bar{v} }$ taken individually have
the form of the JIMWLK-Hamiltonian: $H_v$ is the evolution Hamiltonian for the
dipole operator of the forward amplitude $\hat{N}^0_{\bm{x y}} =
\tr[1-U_{\bm{x}}U^\dagger_{\bm{y}}]/N_c$, which, via the optical theorem,
determines the evolution of the total cross-section.  Real contributions only
occur outside the gap, as mandated by the factor $u(k)$.  If we remove that
restriction by setting $u(k)=1$, we expect complete cancellation of final
state contributions and again a reduction to JIMWLK.  Indeed, setting $u(k)$
to 1 and acting with~(\ref{eq:full-H}) on the dipole operator $\hat{N}_{\bm{x
    y}}^F =\tr [1- (U\bar{U}^\dagger)_{\bm{x}}
(U\bar{U}^\dagger)_{\bm{y}}^\dagger ]/N_c$ (which depends only on products
$(U\bar{U}^\dagger)^{(\dagger)}$), we find that the evolution Hamiltonian
\eqref{eq:full-H} reduces to the JIMWLK-Hamiltonian for Wilson lines
$(U\bar{U}^\dagger)$. The average over $\hat{N}^F_{\bm{x y}}$ then is written
in terms of $\hat{Z}[U, \bar{U}] = \hat{Z}[U\bar{U}^\dagger]$ with evolution
according to
\begin{equation}
    \label{eq:uubar-evolv}
    \partial_\y  \hat{Z}[U\bar{U}^\dagger] 
=
-H [U\bar{U}^\dagger]_{\text{JIMWLK}}
 \hat{Z}[U\bar{U}^\dagger]
\ ,
\end{equation}
cancellation is complete.

The operators that replace $\tr [1- U_{\bm{x}}{U}^\dagger_{\bm{y}}]/N_c$ (i.e.
$\Hat N^F$) in~(\ref{equ-1}) for diffractive dissociation of a photon are
different in- and outside the gap, the structures closely resemble the
factorized results of~\cite{kl}. In the gap, no colored object enters the
final state, thus the initial $q\Bar q$, if in the gap, interacts with the
target and emerges in a singlet state: here $\Hat N^D_{\bm{x y }}$, the
operator for cross sections with rapidity gaps larger than $\y_{\text{gap}}$,
is a product of two traces:
\begin{align}
  \label{eq:ND_ingap}
       \hat{N}^D_{\bm{x y}} =
\hat{ N}^0_{\bm{x y}} \cdot \Hat{\Bar{N}}^0_{\bm{y x}} =  \frac{\tr(
  1-U_{\bm{x}}U^\dag_{\bm{y}} )}{N_c} \frac{ \tr(
  1-\bar{U}^\dag_{\bm{x}}\bar{U}_{\bm{y}} )}{N_c}
\ ,
\end{align}
it takes the form of the ``square'' of two ``elastic'' dipole operators
corresponding to the two amplitude factors.  Here, evolution of amplitude and
complex conjugate are completely uncorrelated ($[H_v, H_{\Bar v}]=0$, $H_r$ is
absent):
\begin{equation}
  \label{eq:gapevo}
  \hat{Z}_\y[U,\Bar U] = e^{-(H_v+H_{\Bar v})(\y-\y_0)} 
  \Hat Z_{\y_0}[U,\Bar U]
\ ,
\end{equation}
and one may factorize $\Hat Z_\y[U,\Bar U]\to
\hat{Z}_\y[U]\hat{Z}_\y[\bar{U}]$ unless the initial condition contradicts 
this.

For $\y > \y_{\text{gap}}$, production of gluons in the final state is allowed
and the $q\bar{q}$-pair can appear also in an octet state. Adding the
octet part to~(\ref{eq:ND_ingap}) removes one trace:
\begin{align}
  \label{eq:ND_outgap}
   \hat{N}_{\bm{x y}}^D 
&= 
   \tr[ 
   \big(1-U^\dag_{\bm{y}}U_{\bm{x}}\big)
   \big(1-\bar{U}^\dag_{\bm{x}}\bar{U}_{\bm{y}} \big)
    ]/{N_c}  
\notag \\
&= 
   \Hat{N}_{\bm{x y}}^0  + \Hat{\Bar{N}}_{\bm{y x}}^0 
   -  \Hat{N}^F_{\bm{x y}} 
\ .
\end{align}
The second line exposes further structure: With the initial conditions on
evolution for $\langle \Hat N^D\rangle$ imposed by~(\ref{eq:ND_ingap}), we
find that $\langle\Hat N^F\rangle$ acquires the interpretation of the cross
section of events with rapidity gaps smaller than $\y_{\text{gap}}$.
Following our previous reasoning we conclude that the average over the three
operators in the second line of \eqref{eq:ND_outgap} can be described by using
functionals $\hat{Z}[U]$, $\hat{Z}[\bar{U}]$ and $\hat{Z}[U\bar{U}^\dagger]$,
with their respective evolution given by a JIMWLK-Hamiltonian. Even if
additional structure in the initial conditions does not prevent these
simplifications, initial conditions for the individual terms are different
from each other (c.f.~(\ref{eq:ND_ingap})) and the inclusive case.

The relation to the results of Kovchegov and Levin~\cite{kl} parallels the
reduction step from JIMWLK to BK: There one observes that JIMWLK evolution
of $\hat{S}_{\bm{x y}}[U] = 1- \hat{N}_{\bm{x y}} = \tr(U_{\bm{x}}
U^\dagger_{\bm{y}})/N_c$ takes the simple form
\begin{align}
  \label{eq:correl_evol}
  \partial_\y \langle \hat{S}_{\bm{x y}}\rangle_\y = \frac{\alpha_s
    N_c}{2\pi^2} \int \!\!d^2 \bm{z} \tilde{\mathcal{K}}_{\bm{x z y}} \langle
  \hat{S}_{\bm{x z}} \hat{S}_{\bm{z y}} - \hat{S}_{\bm{x y}}\rangle_\y
\ ,
\end{align}
where $\tilde{\mathcal{K}}_{\bm{x z y}} = \frac{(\bm{x}-\bm{y})^2}{(\bm{x} -
  \bm{z})^2(\bm{z} - \bm{y})^2} $.  Factorizing $\langle \hat{S}_{\bm{x z}}
\hat{S}_{\bm{z y}} \rangle \to \langle \hat{S}_{\bm{x z}}\rangle \langle
\hat{S}_{\bm{z y}} \rangle$ truncates the infinite Balitsky hierarchy and
leaves us with the BK equation.  For $\y > \y_{\text{gap}}$ where
$\hat{N}^D_{\bm{x y}} = 1 - \hat{S}_{\bm{x y}}[U] - \hat{S}_{\bm{ y
    x}}[\bar{U}] +\hat{S}_{\bm{x y}}[U\bar{U}^\dagger]$ and each
$\hat{S}_{\bm{x y}}[..]$ obeys~\eqref{eq:correl_evol}, the same reasoning
leads to the evolution equation presented as Eq. (9) in~\cite{kl}.
\eqref{eq:ND_ingap} and \eqref{eq:gapevo} imply the required initial condition
$\langle \Hat N^D_{\bm{x y}}\rangle(\y_{\text{gap}}) =\langle \Hat N^0_{\bm{x
    y}}\rangle^2(\y_{\text{gap}})$ for evolution above the gap.

To summarize: We have developed a method which allows to generalize the JIMWLK
approach to a large class of exclusive observables, by simply adapting the
phase space constraints. We have worked out the example of diffractive
dissociation. For all generalizations it is crucial to start from IR safe
observables, otherwise reconstruction of virtual contribution via real virtual
cancellations must fail.

\begin{acknowledgments}
This work was supported in part by BMBF.
\end{acknowledgments}

\end{document}